\documentclass[12pt,preprint]{aastex} 

\newcommand{\ms}{\mbox{m\,s$^{-1}$}}

\newcommand{\msun}{M$_{\odot}$}
\newcommand{\rsun}{R$_{\odot}$}
\newcommand{\mjup}{M$_{\rm JUP}$}

\newcommand{\msini}{$m \sin i$}

\shortauthors{Bailey {\it et~al.\/}}
\shorttitle{GJ\,832b: A Jupiter-like Planet}
\slugcomment{ApJ   Submitted:                 Accepted:             }
\begin{document}

\title{A Jupiter--like Planet Orbiting the Nearby M Dwarf GJ 832\altaffilmark{1}}

\author{Jeremy Bailey\altaffilmark{2},
R. Paul Butler\altaffilmark{3}, 
C. G. Tinney\altaffilmark{4},
Hugh R. A. Jones\altaffilmark{5},
Simon O'Toole\altaffilmark{5,6},
Brad D. Carter\altaffilmark{7},
Geoffrey W. Marcy\altaffilmark{8}
}

\email{jbailey@els.mq.edu.au}

\altaffiltext{1}{Based on observations obtained at 
the Anglo--Australian Telescope, Siding Spring, Australia.}

\altaffiltext{2}{Department of Physics, Macquarie University,
NSW 2109, Australia}

\altaffiltext{3}{Department of Terrestrial Magnetism,
Carnegie Institution of Washington, 5241 Broad Branch Road NW,
Washington D.C. USA 20015-1305}

\altaffiltext{4}{Department of Astrophysics, School
of Physics, University of New South Wales, NSW 2052,
Australia}

\altaffiltext{5}{Centre for Astrophysical Research,
University of Hertfordshire, Hatfield, AL10 9AB, UK}

\altaffiltext{6}{Anglo--Australian Observatory, P.O. Box 296,
Epping, NSW 1710, Australia}

\altaffiltext{7}{Faculty of Sciences, University of Southern
Queensland, Toowoomba, Queensland 4350, Australia}

\altaffiltext{8}{Department of Astronomy, University of California, Berkeley, CA
94720.}

\begin{abstract}

Precision Doppler velocity measurements from the 
Anglo-Australian Telescope reveal a planet with a 9.4$\pm$0.4 year
period orbiting the M1.5 dwarf GJ\,832.  Within measurement
uncertainty the orbit is circular, and the minimum mass (\msini)
of the planet is 0.64$\pm$0.06\,\mjup.  GJ\,832 appears to be depleted in metals by 
at least 50\% relative to the Sun, as are a significant fraction of
the M dwarfs known to host exoplanets. GJ\,832 adds another Jupiter-mass
planet to the known census of M dwarf exoplanets, which currently includes 
a significant number of Neptune-mass planets. GJ\,832 is an excellent
candidate for astrometric orbit determination with $\alpha \sin i = 0.95$ mas.
GJ\,832b has the second largest angular distance from its star among radial
velocity detected exoplanets (0.69 arc sec) making it a potentially interesting
target for future direct detection.

\end{abstract}

\keywords{planetary systems -- stars: individual (GJ 832)}

\section{Introduction}
\label{intro}

Most of the known exoplanets orbit late-F, G, or early-K dwarfs,
with masses ranging from 0.7 to 1.2\,\msun.  There are nearly
2,000 such stars within 50\,pc brighter than V=8.  In contrast
there are just a handful of M dwarfs brighter than V=8. 

Although M dwarfs make up 70\% of nearby stars, their faintness in
the optical makes them  difficult targets
for which to obtain precision Doppler velocities. As a result 
they make up as little as 5\% of the current planet search targets, and
only 11 exoplanets
have been found to date, orbiting a total of 7 M dwarfs.  
Just over half
of these M dwarf planets are less massive than Neptune, leading to
speculation that M dwarfs typically host either fewer planets than G dwarfs 
\citep{johnson07}, or lower mass planets as a result of their
smaller proto-planetary disks \citep{laughlin04,il05,kk08}.

The primary parameters in planet formation theory are the
mass of the central star, the mass of the protoplanetary
disk, and the metallicity of the system.  While it is now
well established for late-F, G, and K dwarfs that metal-rich stars are
enhanced in planets relative to metal-poor stars \citep{g1,g2,g4,g5,g3,sb,s1,s04,g6,re,fv,bond}, 
it has been harder to establish
the importance of stellar mass on planet formation since most
of the stars under survey lie in the relatively narrow
mass range encompassed by late-F, G and K dwarfs.

The searches which {\em are} being done for planets orbiting M dwarfs,
will ultimately provide the data needed to see if
the metallicity-planet relation extends down to the M dwarf
regime, and whether the mass distribution of exoplanets formed around
M dwarfs is similar to, or different than, that for more massive host stars.
M dwarf Doppler surveys, therefore,  have the power to address some of the most important 
questions in exoplanetary science, as they extend the
mass range of potential exoplanet host stars down to 0.3\,\msun.  

We report here a new extrasolar planet in a long period orbit with
eccentricity consistent with zero, discovered by the Anglo-Australian
Planet Search (AAPS).  The AAPS program is
described in Section 2.  The characteristics of the host star
and our Doppler measurements are presented in Section 3.
A discussion follows.

\section{The Anglo-Australian Planet Search}

The AAPS began in 1998 January, and is
currently surveying 250 stars.  Thirty exoplanets with
\msini\ ranging from 0.17 to 10\,\mjup\ have first been discovered
by the AAPS \citep{AAPSI,AAPSIII,AAPSVII,AAPSXI,AAPSXIII,AAPSII,AAPSV,
AAPSIV,AAPSVI,AAPSVIII,AAPSXII,AAPSIX,AAPSX,AAPSXIV}.
Our precision Doppler measurements are made with the UCLES echelle
spectrometer \citep{Diego90} on the 3.9m Anglo-Australian Telescope (AAT).  
An iodine absorption cell provides wavelength calibration from
5000 to 6200 \AA.  The spectrometer point-spread function and wavelength calibration
are derived from the iodine absoption lines embedded on every
spectrum by the cell \citep{Valenti95,Butler96}.
Our observing and analysis  system has demonstrated long term
precision of 3\,\ms\ for
late-F, G, and early-K dwarfs brighter than V=7.5 \citep{AAPSXI,AAPSII}.

\section{GJ 832}
\label{obs}
 
At 4.93\,pc, GJ\,832 (LHS\,3865, HD\,204961, HIP\,106440) is amongst the nearest
stars in the sky \citep{Perryman97}.  It is an M1.5 dwarf 
with an optical absolute magnitude and colors of M$_{\rm V}=10.19$, $V=8.66$, 
and $B-V=1.52$, and an infrared absolute magnitude and colors of M$_{\rm K}=6.03$, $K=4.50$, 
and $V-K=4.16$.
Both Hipparcos and ground-based photometry \citep{Koen02}
find GJ 832 to be photometrically stable at the several milli-magnitude
level.  
\citet{Gautier07} have combined Nstars\footnote{http://nstars.nau.edu} 
visible photometry with Spitzer far-infrared photometry, to
estimate an ``infrared flux method'' effective temperature of 3657\,K for GJ\,832.  The Spitzer
observations reveal no evidence of mid- or far-infrared excess.  
The radius of GJ\,832 is estimated to be 0.48\,\rsun\ \citep{Pasinetti01}.

Although accurate metallicities for M dwarfs are problematic, 
GJ\,832 is likely to be rather metal poor.  Matching synthetic spectra to high-resolution 
spectra of the FeH band near 9900\AA,
Schiavon et al. (1997) estimate a metallicity for GJ\,832
of [Fe/H] $=$ $-0.7$, and a surface gravity of $\log g = 4.7$.
The photometric metallicity calibration of \citet{Bonfils05a}
gives an estimated metallicity of [Fe/H] = $-0.31\pm0.2$.

Due to its late spectral type, GJ\,832 has not (to date) been subject to detailed 
spectroscopic analysis, and so to estimate its mass we must rely on either
theoretical isochrones, or empirical mass-luminosity calibration.
The latter indicate a mass for GJ\,832 of 0.45$\pm$0.05\,\msun\ (with the mass uncertainty
being largely due to the scatter about the mass-luminosity calibration relationship
of \citet{delfosse98}). The Padova theoretical isochrones \citep{marigo08} predict
M$_{\rm K}$ ranging from 5.97 (at $10^9$\,yr) to 6.03 (at $10^{10}$\,yr) for a 0.45\,\msun\
dwarf with [Fe/H] = $-0.3$ which is consistent with the observed luminosity of GJ\,832.
At [Fe/H] = $-0.7$ they predict M$_{\rm K}$ in the range 5.92 (at $10^9$\,yr) to 
5.86 (at $10^{10}$\,yr). Given the difficulty in determining metallicities for M dwarfs,
we therefore derive a mass estimate for the primary of 0.45$\pm$0.05\,\msun.

\begin{figure}
\epsscale{.8}
\plotone{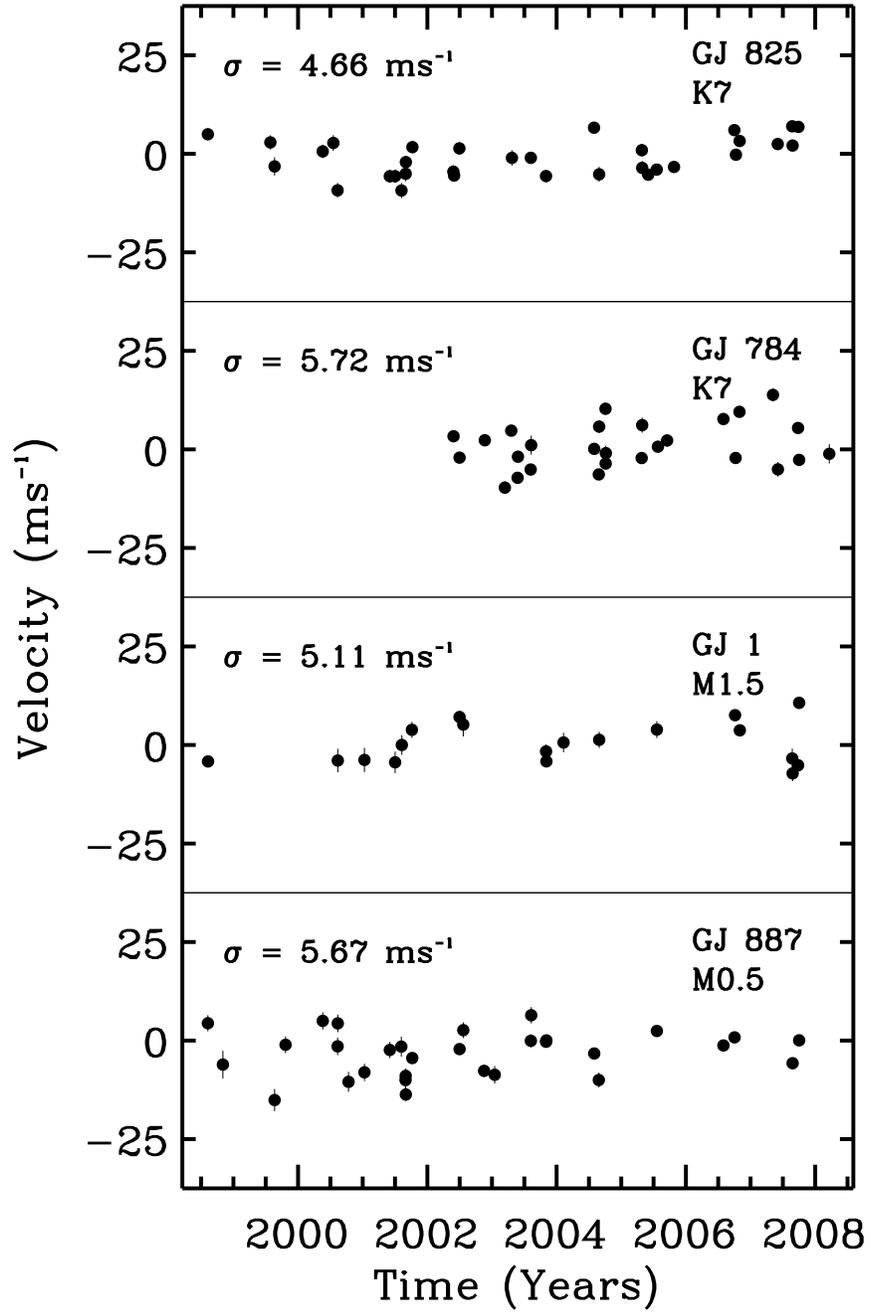}
%\centerline{\scalebox{.75}{\rotatebox{90}{\includegraphics{fig2.ps}}}}
\caption{Four stable late K and M dwarfs from the AAT.}
\label{stable_stars}
\end{figure}

GJ\,832 is chromospherically quiescent.  Based on high
resolution spectroscopy of the CaII H\&K lines, \citet{CaHKI}
report $log R^\prime_{\rm HK} = -5.10$.  
This would suggest a jitter of 3.9\,\ms\ using the
$B-V$, M$_{\rm V}$ \& T$_{\rm eff}$ in the most recent stellar ``jitter''
calibration of J.Wright (priv.comm).
\citet{Bonfils05b} 
estimate the stellar jitter of GJ\,832 to be less than 2\,\ms.
GJ\,832 is among the fainter stars on the AAT program.  The
signal-to-noise of these observations range from 46 to 150 per spectral pixel, with
a median of 98, which is lower than typical for AAPS targets.  Four
late dwarfs from the long term AAT program are shown in
Figure \ref{stable_stars}.  These stars are shown in order of descending B$-$V.
GJ 887 is an especially close match to GJ\,832 in B-V colour 
and V magnitude.  Based on this we estimate the combined velocity uncertainty due
to photon statistics, jitter, unknown planets,
and systematic errors is 5\,\ms\ for late-K and M dwarfs
in the AAPS.  This is comparable to that estimated for late-K and M
dwarfs in the Keck program, as shown in Figures 2-4 of \citet{Butler08}.

\begin{figure}
\epsscale{1.0}
\plotone{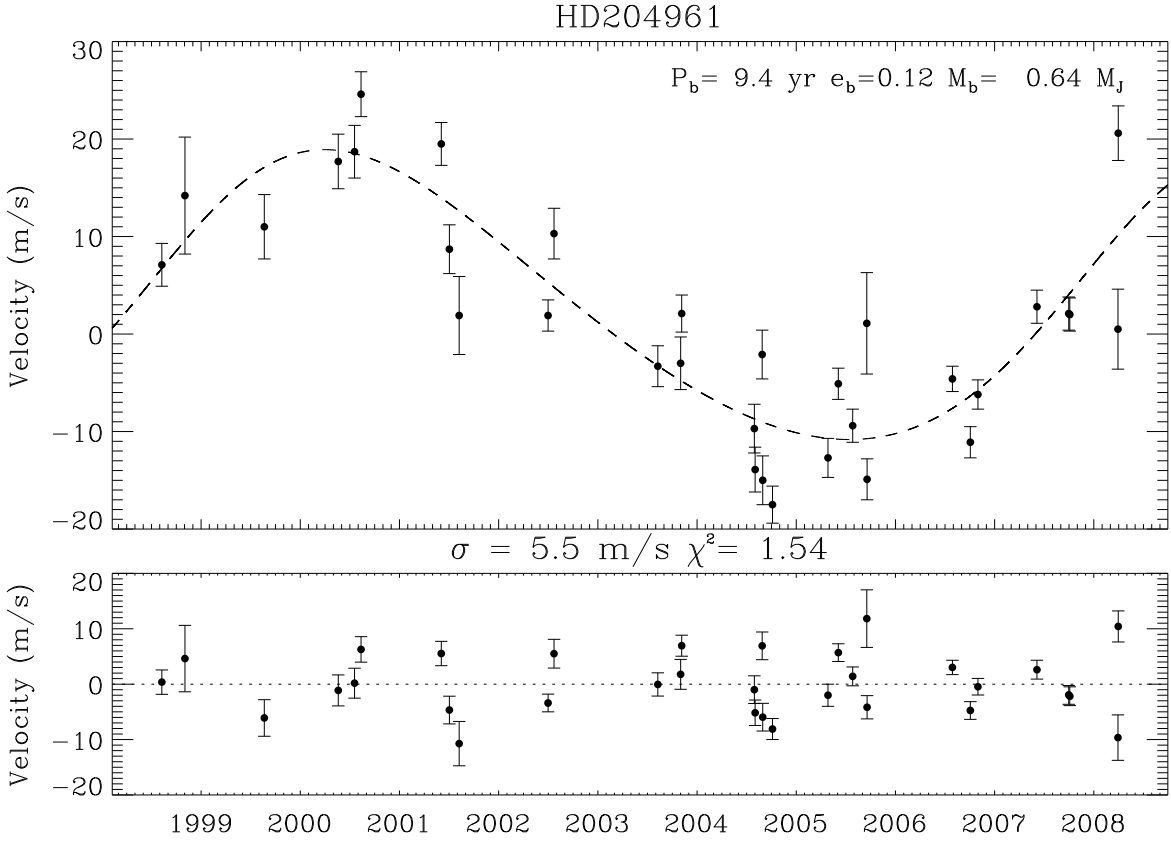}
%\centerline{\scalebox{.75}{\rotatebox{90}{\includegraphics{fig2.ps}}}}
\caption{Doppler velocities for GJ\,832 spanning 9.6 yr.
The upper panel shows the measured velocities with a best-fit Keplerian over-plotted as
a dashed line. The residuals to this fit are plotted in the
lower panel. The Keplerian orbital parameters obtained listed in see Table \ref{orbit},
and strongly suggest the presence of a \msini=0.64\,\mjup\ exoplanet.}
\label{GJ832_Velocities}
\end{figure}

A total of 32 precision Doppler measurements of GJ\,832
spanning 9.6 years are listed in Table \ref{velgj832} and shown in
Figure \ref{GJ832_Velocities} (upper panel).  The root-mean-square (RMS) scatter
of the residuals about the mean velocity of this data set is 11.6\,\ms.Using the
2-Dimensional Keplerian Lomb-Scargle (2DKLS) peridogram of \citet{AAPSXIV} to
identify an initial period and eccentricity, the subsequent
best-fit Keplerian to all 32 epochs of data reduces this to an RMS of 5.5\,\ms,
and gives a reduced $\chi_{\nu}^2$ of 1.54 (see Table \ref{orbit} -- a stellar jitter
of 3.9\,\ms\ was used, together with the internal velocity measurement uncertainty
for each epoch in Table \ref{velgj832}, to determine reduced $\chi_{\nu}^2$).
These fit parameters strongly suggest the presence of an exoplanet with minimum
mass \msini\ of 0.64\,\msun, period 9.4$\pm$0.4\,yr, eccentricity 0.12$\pm$0.11 (which we consider to be consistent with
zero eccentricity, particularly when the bias against measuring zero eccentricities
demonstrated by \citet{otoole08} is taken into account) and semi-major axis 3.4$\pm$0.4\,AU.

\begin{figure}
\includegraphics[angle=-90,width=\textwidth]{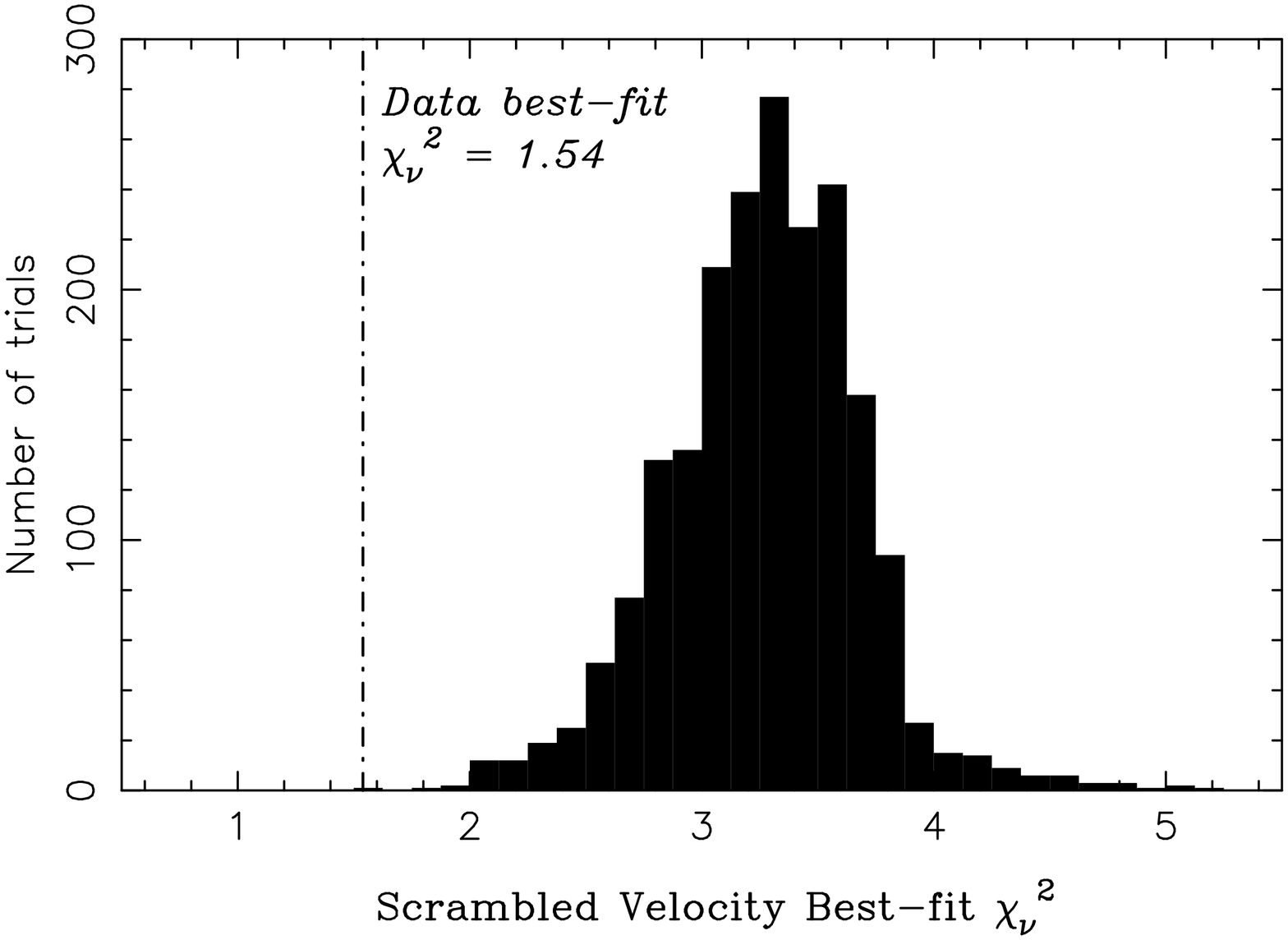}
%\centerline{\scalebox{.75}{\rotatebox{90}{\includegraphics{fig2.ps}}}}
\caption{Assessment of the FAP of the Keplerian model for GJ\,832. The histogram
shows the values of $\chi_{\nu}^2$ from 2002 trials with randomly scrambled
velocities. Only one of these trials had $\chi_{\nu}^2$ lower than the value of
1.54 from the original fit, implying a FAP of 0.05\%}
\label{FAP}
\end{figure}

We have determined the False Alarm Probability (FAP, i.e. the probability
that we have falsely identified an exoplanet that is not present) for this orbit
determination using the Monte Carlo ``scrambled velocities'' approach described by
\cite{marcy05}. This method tests the hypothesis that no planet is present and
the Keplerian fit could have been obtained from mere noise, by generating randomly scrambled data sets in which
the order of velocities are changed but the times remain the same. These are
then subjected to the same analysis as our actual data set (i.e. identfying the
strongest peak in the 2DKLS followed by a full Keplerian fit). In this case
2002 random trials were carried out and only one of these yielded a
$\chi_{\nu}^2$ less than the value of 1.54 obtained with the original data. The
histogram of $\chi_{\nu}^2$ is shown in Figure \ref{FAP}. These results
imply a FAP of 0.05\% for the GJ 832 planet detection.

\section{Discussion}

GJ\,832 at a distance of 4.93pc is one of the nearest known exoplanetary
systems. The combination of the small distance and relatively long period 
gives a large angular distance from the star
of 0.69 arc seconds for an edge-on circular orbit. This is exceeded only by
$\epsilon$ Eri among radial velocity detected exoplanets, and only six other
systems exceed 0.2 arc sec. GJ\,832b is therefore a potentially interesting
target for direct detection, although the high constrast with the star 
\citep[likely to be $<10^{-8}$;][]{Burrows04} still makes this an extremely 
challenging observation.

GJ\,832 is an excellent candidate for astrometric orbit determination. The
astrometric orbit semimajor axis is $\alpha \sin i = 0.95$ mas, which is
comparable to that of $\epsilon$ Eri for which an astrometric orbit was
determined by \cite{benedict06} and larger  than that of GJ 876 
which also has an astrometric orbit determination \citep{benedict02}. The
astrometric orbit would enable the inclination to be determined, removing the
current $\sin i$ uncertainty on the mass. 

Seven M dwarfs (including GJ\,832) are currently known to host as many as 
11 exoplanets, and
these are listed in Table \ref{mdwarfs} (see table notes for
references).  
As noted earlier, determining the metallicities
of M dwarfs is notoriously difficult -- published metallicity estimates are
available for several of the known exoplanet host M dwarfs, and these
are listed in the Table. In addition, we have also derived for all seven
M-dwarfs a photometric
metallicity estimate, using  the technique of \citet{Bonfils05a},
which has the advantage of being uniform over all these M dwarfs.
On average the \citet{Schiavon97} metallicities appear to be systematically
0.3-0.4\,dex lower than those derived from the Bonfils et al. calibration.
The \citet{bean06} metallicities are similarly on the metal-poor side 
of the Bonfils et al. results, though not by as much ($\approx$ 0.2\,dex).
In general, the metallicity trends are similar across all three calibrations,
and it is clear there is a metallicity spread across the observed M dwarf exoplanet
hosts.

Based on these metallicity estimates, it would appear that four of
the current M dwarf exoplanet host stars are somewhat metal-poor, two have about
solar metallicity, and one is slightly metal rich.
Given the well known correlation between stellar
metallicity and observed exoplanet frequency for F, G, and K dwarf host stars, 
this metallicity distribution for M dwarf host stars
is quite unexpected. Whhile the numbers of systems are small there is no obvious
difference in metallicity between the stars hosting Jupiter-mass planets and
those hosting Neptune-mass planets.

The correlation between high stellar metallicity and
planets for late-F, G, and early-K dwarfs points
toward the core accretion model for planet formation.
But there does not appear to be strong evidence to date that
M dwarf planet formation is strongly correlated with
high metallicity. This is puzzling, particularly in view of the fact that
M dwarfs probably have lower mass protoplanetary disks, and therefore would
need even higher metallicity than a F, G or K dwarf to provide enough solid
material (silicates and ices) to build a planetary core. Obviously it must be 
kept in mind that measuring metallicities for
M dwarfs is problematic, and that even the \citet{Bonfils05a}
calibration (though empirically based and moderately robust) is only
good to $\pm$0.2\,dex. Nonetheless it is interesting to
consider possible means by which M dwarf exoplanets could be formed in such a 
manner as to {\em not} display the strong metallicity correlation seen in
FGK dwarfs. One initially attractive explaneation is that since  M dwarfs are essentially
immortal on a Hubble time scale, the vast majority of nearby M dwarfs could be
old metal poor stars. Unfortunately such an explanation
would appear unlikely. The study of M dwarf
kinematics has an extensive and venerable history \citep[e.g.][]{wielen77, wu95, rhg95}
which has contributed to the creation of extensive and sophisticated models of the
stellar populations present in the Solar Neighbourhood \citep[e.g. the Besancon models
of][]{robin2003}.
More recently, the availability of huge numbers of M dwarf spectra from the
SDSS survey have enabled sophisticated tests of the kinematics of the Besancon models
by \citet{bochanski07}, and  have substantially born out the
Besancon model predictions for M dwarfs. Those models indicate that 
dominant solar neighbourhood M dwarfs will be thin disc members with ages almost
uniformly spread between 0.1 and 10\,Gyr. Thick disc M dwarfs (which would indeed
be expected to have systematically lower metallicities)
will be present at much lower densities \citep[around a factor of one twentieth or 
less;][]{robin2003}, and the probability that they would make up four of the
seven M dwarf exoplanet hosts would seem to be negligibly small.
 
An alternative explanation could be that M dwarf planets might form primarily via
the disk instability mechanism \citep[see e.g.][and references therein]{boss08}, 
rather than via core accretion,
which would make their formation probability more or less independent of metallicity.

Six of the eleven exoplanets known to orbit M dwarfs have
minimum masses less than 0.1\,\mjup.  In contrast,
only nine planets with \msini $<$ 0.1\,\mjup\ have been
found among the 216 Doppler velocity planets with
B-V $<$ 1.2 in the ``Catalog of Nearby Exoplanets'' \citep{butler06b}.  
With a minimum mass of
0.64$\pm$0.06\,\mjup, GJ\,832b is the fifth jovian mass planet
found orbiting an M dwarf.  The most massive M dwarf
planet yet found is 1.93\,\mjup.  Since massive planets
are by far the easiest ones to find, planets of more than 2\,\mjup\
orbiting within 3\,AU of M dwarfs must be rare,
occuring less than around once per 300 M dwarfs.

\acknowledgements

We acknowledge support by NSF grant AST-9988087, NASA grant
NAG5-12182, PPARC grant PP/C000552/1, ARC Grant DP0774000
and travel support from the Carnegie Institution
of Washington (to RPB) and from the Anglo-Australian Observatory (to CGT, BDC
and JB).  
We thank the Anglo-Australian time assignment
committees for allocations of AAT time.  We are grateful
for the extraordinary support we have received from the
AAT technical staff  -- E. Penny, R. Paterson, D. Stafford,
F. Freeman, S. Lee, J. Pogson, S. James, J. Stevenson, K. Fiegert and G. Schaffer.

\clearpage

\begin{deluxetable}{rrr}
\tablecaption{Velocities for GJ\,832\label{velgj832}}

\tablewidth{0pt}
\tablehead{
\colhead{JD}           &    \colhead{RV}         & \colhead{Uncertainty} \\
\colhead{($-$2451000)}   &  \colhead{(m\,s$^{-1}$)} & \colhead{(m\,s$^{-1}$)} 
}
\startdata
    34.0873  &     6.8  &  2.2 \\
   119.0159  &    14.0  &  6.0 \\
   411.1222  &    10.8  &  3.3 \\
   683.2628  &    17.4  &  2.8 \\
   743.1456  &    18.4  &  2.7 \\
   767.0812  &    24.4  &  2.3 \\
  1062.2443  &    19.2  &  2.2 \\
  1092.1677  &     8.4  &  2.5 \\
  1128.1273  &     1.6  &  4.0 \\
  1455.2341  &     1.7  &  1.6 \\
  1477.1455  &    10.0  &  2.6 \\
  1859.0874  &    -3.5  &  2.1 \\
  1943.0361  &    -3.3  &  2.7 \\
  1946.9712  &     1.8  &  1.9 \\
  2214.2066  &   -10.0  &  2.5 \\
  2217.2117  &   -14.2  &  2.3 \\
  2243.0503  &    12.8  &  2.5 \\
  2245.1511  &   -15.4  &  2.5 \\
  2281.0469  &   -17.7  &  1.9 \\
  2485.3011  &   -12.9  &  2.0 \\
  2523.3005  &    -5.3  &  1.6 \\
  2576.1420  &    -9.7  &  1.7 \\
  2628.0699  &     1.0  &  5.2 \\
  2629.0549  &   -15.2  &  2.1 \\
  2943.1074  &    -4.8  &  1.3 \\
  3009.0378  &   -11.3  &  1.6 \\
  3036.9559  &    -6.5  &  1.5 \\
  3254.2003  &     2.7  &  1.7 \\
  3371.0670  &     1.6  &  1.7 \\
  3375.0442  &     2.0  &  1.7 \\
  3552.2912  &     6.8  &  4.1 \\
  3553.3041  &    17.2  &  2.8 \\
\enddata
\end{deluxetable}

\begin{deluxetable}{lc}
\tablecaption{Orbital Solutions for GJ\,832\label{orbit}}
%
% Simon O'Toole's fits of 01 July 2008.
%
\tablewidth{0pt}
\tablehead{
\colhead{Parameter} & \colhead{Value} 
}
\startdata
Orbital period $P$ (days)                &  3416$\pm$131 \\
Velocity semiamplitude $K$ (\ms)         &  14.9$\pm$1.3 \\
Eccentricity $e$                         & 0.12$\pm$0.11 \\
Periastron date (Julian Date$-$2451000)  & 211$\pm$353   \\
$\omega$ (degrees)                       & 304$\pm$38    \\
$m \sin i$ (\mjup)                        & 0.64$\pm$0.06 \\
semimajor axis $a \sin i$ (AU)           & 3.4$\pm$0.4   \\ 
N$_{\rm obs}$                            & 32            \\
RMS (m\,s$^{-1}$)                        & 5.5           \\
$\chi_\nu^2$                             & 1.54          \\
\enddata
\end{deluxetable}

\begin{deluxetable}{llrrrrrll}
\tablecaption{Known M dwarf Exoplanet Hosts\label{mdwarfs}}

\tablewidth{0pt}
\tablehead{
\colhead{Host} 
       & \colhead{Type}
                &\colhead{M$_{\rm K}$}
		     & \colhead{$V-K$}
                             & \colhead{Phot [Fe/H]\tablenotemark{a}} 
                                     & \colhead{[Fe/H]\tablenotemark{b}} 
                                     & \colhead{[Fe/H]\tablenotemark{c}} 
                                          & \colhead{\msini(\mjup)} 
                                                              & \colhead{Refs\tablenotemark{d}}
}
\startdata
                                                                             %          V     K      pi   +-pi                Ksrc
GJ\,832 &  M1.5 & 6.03 & 4.16& -0.31 & -0.7  &       & 0.64              & 1 \\         %GJ 832   8.66  4.50 202.52  1.33                2MASS
GJ\,876 &  M4   & 6.64 & 5.16&  0.02 & -0.4  & -0.12 & 0.019,0.619,1.935 & 2,8,9 \\    %GJ 876  10.17  5.01 212.69  2.10                2MASS
GJ\,849 &  M3.5 & 5.87 & 4.83&  0.16 &       &       & 0.82              & 4 \\         %GJ 849  10.42  5.59 113.97  2.10  5.874  -0.090 2MASS
GJ\,317 &  M3.5 & 7.26 & 4.97& -0.23 &       &       & 1.2               & 7 \\         %GJ 317  13.0   7.03 111     12    7.255  -0.206 2MASS
GJ\,436 &  M2.5 & 6.02 & 4.61& -0.02 &       & -0.32 & 0.067             & 3 \\         %GJ 436  10.68  6.07  97.73  2.27  6.020  -0.281 2MASS
GJ\,581 &  M2.5 & 6.85 & 4.72& -0.26 & -0.1  & -0.33 & 0.049,0.016,0.026 & 5,10 \\      %GJ 581  10.56  5.84 159.52  2.27                2MASS
GJ\,674 &  M3   & 6.57 & 4.50& -0.30 &       &       & 0.035             & 6 \\         %GJ 674   9.36  4.86 220.43  1.63  6.571  -0.577 2MASS

\enddata
\tablenotetext{a}{Photometric [Fe/H] determined using catalogued V, 2MASS K$_{\rm s}$ and 
parallax data, with the \citet{Bonfils05a} relation.}
\tablenotetext{b}{Metallicity estimates from \citet{Schiavon97}}
\tablenotetext{c}{Metallicity estimates from \citet{bean06}}
\tablenotetext{d}{M dwarf exoplanet properties from; 1 - this paper; 2 -
\cite{delfosse98}; 3 - \citet{butler04}; 4 - \citet{butler06a}; 5 - \citet{Bonfils05b};
6 - \citet{Bonfils07}; 7 - \citet{johnson07}; 8 - \citet{marcy01} ; 9 - \citet{rivera05};
10 - \citet{udry07}}

\end{deluxetable}

\clearpage

\clearpage

\begin{thebibliography}{}


\bibitem[\protect\citeauthoryear{Bean et al.}%
  {2006}]{bean06} Bean, J. L., Benedict G. F. \& Endl, M., 2006, ApJ, 653, L65

\bibitem[\protect\citeauthoryear{Benedict et al.}{2002}]{benedict02}
  Benedict, G. F. et al., 2002, ApJ, 581, L115
  
\bibitem[\protect\citeauthoryear{Benedict et al.}{2006}]{benedict06}  
  Benedict, G. F. et al., 2006, AJ, 132, 2206
  
\bibitem[\protect\citeauthoryear{Bond et al.}%
  {2006}]{bond} Bond, J. C., Tinney, C. G., Butler, P. R., Jones, H. R. A., Marcy, G. W.,
  Penny, A. J. \& Carter, B. D., 2006, MNRAS, 370, 163

\bibitem[\protect\citeauthoryear{Bochanski et al.}{2007}]{bochanski07}
   Bochanski, J. J., Munn, J. A., Hawley, S. L., West, A. A.,
   Covey, K. R. \& Schneider, D. P. 2007, AJ, 134, 2418

\bibitem[\protect\citeauthoryear{Bonfils et al.}{2005a}]{Bonfils05a}
   Bonfils, X., Delfosse, X., Udry, S., Santos, N.C., Forveille, T. \& 
   Segransan, D. 2005a, A\&A, 442, 635

\bibitem[\protect\citeauthoryear{Bonfils et al.}{2005b}]{Bonfils05b}
   Bonfils, X., et al. 2005b, A\&A, 443, L15

\bibitem[\protect\citeauthoryear{Bonfils et al.}{2007}]{Bonfils07}
   Bonfils, X., et al. 2007, A\&A, 474, 293

\bibitem[\protect\citeauthoryear{Boss}%
   {2008}]{boss08} Boss, A. P. 2008, ApJ, 677, 607
 
\bibitem[\protect\citeauthoryear{Burrows et al.}{2004}]{Burrows04}
   Burrows, A., Sudarsky, D. \& Hubeny, I., 2004, ApJ, 609, 407

\bibitem[\protect\citeauthoryear{Butler et al.}{1996}]{Butler96}
   Butler, R.~P., Marcy, G.~W., Williams, E., McCarthy, C.,
   Dosanjh, P., \& Vogt, S.~S. 1996,
   \newblock { PASP, } {108}, 500

\bibitem[\protect\citeauthoryear{Butler et al.}{2001}]{AAPSII}
   Butler, R.~P., Tinney, C.~G., Marcy, G.~W., Jones, H.~R.~A., Penny, A.~J. \& Apps, K. 2001, 
   \newblock { ApJ }, 555, 410.

\bibitem[\protect\citeauthoryear{Butler et al.}{2002}]{AAPSV}
   Butler, R.P. et al.  2002, \apj, 578, 565 %more than 8 authors

\bibitem[\protect\citeauthoryear{Butler et al.}{2004}]{butler04}
   Butler, R. P., Vogt, S. S., Marcy, G. W., Fischer, D. A., Wright, J. T., Henry, G. W., Laughlin, G., \& Lissauer, J. J. 2004, ApJ, 617, 580

\bibitem[\protect\citeauthoryear{Butler et al.}{2006a}]{butler06a}
   Butler, R. P., Johnson, J. A., Marcy, G. W., Wright, J. T., 
   Vogt, S. S. \& Fischer, D. A. 2006a, PASP, 118, 1685

\bibitem[\protect\citeauthoryear{Butler et al.}{2006b}]{butler06b}
   Butler, R. P., et al. 2006b, ApJ, 646, 505

\bibitem[\protect\citeauthoryear{Butler et al.}{2008}]{Butler08}
   Butler, R. P., Howard, A. W., Vogt, S.S. \& Wright, J. T. 2008, ApJ, submitted 

\bibitem[\protect\citeauthoryear{Carter et al.}{2003}]{AAPSIX}
  Carter, B. D., Butler, R. P., Tinney, C. G., Jones, H. R. A., Marcy, G. W., McCarthy,
  C., Fischer, D. A. \& Penny,A. J. 2003, ApJ, 593, L43

\bibitem[\protect\citeauthoryear{Delfosse et al.}{1998}]{delfosse98}
   Delfosse, X., Forveille, T., Mayor, M., Perrier, C., Naef, D., \& Queloz, D. 1998, A\&A, 338, L67

\bibitem[\protect\citeauthoryear{Diego et~al.}{1990}]{Diego90}
   Diego, F., Charalambous, A., Fish, A.~C., \& Walker, D.~D. 1990, Proc. Soc. Photo-Opt. Instr. Eng., 1235, 562

\bibitem[\protect\citeauthoryear{Fischer \& Valenti}%
    {2005}]{fv} Fischer, D. A. \& Valenti, J., 2005,
    ApJ, 662, 1102

\bibitem[\protect\citeauthoryear{Gautier et~al.}{2007}]{Gautier07}
   Gautier III, T.~N., Rieke, G.~H., Stansberry, J., Byrden, G.~C., Stapelfeldt, K.~R., Werner, M.~W., Beichman, C.~A., Chen, C., Su, K., Trilling, D., Patten, B.~M., \& Roellig, T.~L. 2007,
   \newblock {  ApJ, }, 667, 527.

\bibitem[\protect\citeauthoryear{Gonzalez}{1997}]{g1} 
  Gonzalez, G., 1997, MNRAS, 285, 403
\bibitem[\protect\citeauthoryear{Gonzalez}{1998}]{g2} 
  Gonzalez, G., 1998, A\&A, 334, 221
\bibitem[\protect\citeauthoryear{Gonzalez \& Vanture} {1998}]{g4} 
  Gonzalez, G. \& Vanture, A. D., 1998,
  A \& A, 339, L29
\bibitem[\protect\citeauthoryear{Gonzalez \& Laws} {2000}]{g3} 
  Gonzalez, G. \& Laws, C., 2000, AJ, 119, 390
\bibitem[\protect\citeauthoryear{Gonzalez et al.} {1999}]{g5} 
  Gonzalez, G., Wallerstein, G. \& Saar, S. H., 1999,
  ApJ, 511, L111
\bibitem[\protect\citeauthoryear{Gonzalez et al.} {2001}]{g6} 
  Gonzalez, G., Laws, C., Tyagi, S. \& Reddy, B. E., 2001,
  AJ, 121, 432
    
\bibitem[\protect\citeauthoryear{Ida \& Lin} {2005}]{il05}
   {Ida}, S. \& {Lin}, D.~N.~C. 2005,
   \newblock {  ApJ, }, 626, 1045.

\bibitem[\protect\citeauthoryear{Johnson et al.} {2007}]{johnson07}
   Johnson, J. A., Butler, R. P., Marcy, G. W., Fischer, D. A., Vogt, S. S., Wright, J. T., \& Peek, K. M. G. 2007, ApJ, 670, 833

\bibitem[\protect\citeauthoryear{Jones et al.}{2002}]{AAPSIV} 
   Jones, H. R. A., Butler, R. P., Marcy, G. W., Tinney, C. G., 
   Penny, A. J., McCarthy, C. \& Carter, B. D. 2002, \mnras, 333, 871

\bibitem[\protect\citeauthoryear{Jones et al.}{2003a}]{AAPSVI} 
   Jones, H. R. A., Butler, R. P., Marcy, G. W., Tinney, C. G., 
   Penny, A. J., McCarthy, C. \& Carter, B. D. 2003a, \mnras, 337, 1170

\bibitem[\protect\citeauthoryear{Jones et al.}{2003b}]{AAPSVIII} 
   Jones, H. R. A., Butler, R. P., Marcy, G. W., Tinney, C. G., 
   Penny, A. J., McCarthy, C., Carter, B. D. \& Pourbaix, D. 2003b, \mnras, 341, 948

\bibitem[\protect\citeauthoryear{Jones et al.}{2006}]{AAPSXII} 
   Jones, H. R. A., Butler, R. P., Tinney, C. G., Marcy, G. W.,  
   Carter, B. D., Penny, A. J., McCarthy, C., \& Bailey, J. A. 2006, \mnras, 369, 249

\bibitem[\protect\citeauthoryear{Kennedy \& Kenyon}{2008}]{kk08}
   {Kennedy}, G.~M. \& {Kenyon}, S.~J., 2008
   \newblock {  ApJ, }, 673, 502.

\bibitem[\protect\citeauthoryear{Koen et~al.}{2002}]{Koen02}
   Koen, C., Kilkenny, D., Van Wyk, F.,
   Cooper, D., \& Marang, F. 2002,
   \newblock {  MNRAS, }, 334, 20.

\bibitem[\protect\citeauthoryear{Laughlin et~al.}{2004}]{laughlin04}
   {Laughlin}, G., {Bodenheimer}, P. \& {Adams}, F.~C. 2004,
   \newblock {  ApJ, }, 612, L73.

\bibitem[\protect\citeauthoryear{McCarthy et~al.}{2004}]{AAPSX}
  McCarthy, C. et al, 2004, ApJ, 617, 575 %more than 8 authors
  
\bibitem[\protect\citeauthoryear{Marcy et al.} {2001}]{marcy01}
   Marcy, G. W., Butler, R. P., Fischer, D., Vogt, S. S., Lissauer, J. J, \& Rivera, E. J. 2001, ApJ, 556, 296

\bibitem[\protect\citeauthoryear{Marcy et al.}{2005}]{marcy05}
   Marcy, G. W., Butler, R. P., Vogt, S. S., Fischer, D. A., Henry, G. W., Laughlin,
   G., Wright, J. T. \& Johnson, J. A., 2005, ApJ, 619, 570

\bibitem[\protect\citeauthoryear{Marigo et al.} {2008}]{marigo08}
   Marigo, P., Girardi, L., Bressan, A., Groenewegen, M. A. T., Silva, L. \& Granato, G. L., 2008, A\&A, 482, 883.
   
\bibitem[\protect\citeauthoryear{O'Toole et al.} {2007}]{AAPSXIV}
   O'Toole, S., et al.  2007, ApJ, 660, 1636

\bibitem[\protect\citeauthoryear{O'Toole et al.} {2008}]{otoole08}
   O'Toole, S., Tinney, C. G., Jones, H. R. A., Butler, R. P., Marcy, G. W.,   
   Carter, B. D. \& Bailey, J. A.  2008, MNRAS, submitted.
 

\bibitem[\protect\citeauthoryear{Pasinetti-Fracassini et~al.}{2001}]{Pasinetti01}
   Pasinetti-Fracassini, L.~E., Pastori, L., Covino, S., \& Pozzi A. 2001,
   \newblock { A\&A, } 367, 521.  

\bibitem[\protect\citeauthoryear{Perryman et al.}{1997}]{Perryman97}
   Perryman, M.~A.~C., et al. 1997, { A\&A, } 323, L49.
   The Hipparcos Catalog

\bibitem[\protect\citeauthoryear{Reid}{2002}]{re} 
    Reid, I. N., 2002, PASP, 114, 306

\bibitem[\protect\citeauthoryear{Reid et al.} {1995}]{rhg95}
  Reid, I. N., Hawley, S. L. \& Gizis, J. E. 1995, AJ, 110, 1838.

\bibitem[\protect\citeauthoryear{Rivera et al.}{2005}]{rivera05}
   Rivera, E. J., Lissauer, J. J., Butler, R. P., Marcy, G. W., Vogt, S. S., Fischer, D. A., Brown, T. M., Laughlin, G., \& Henry, G. W. 2005, ApJ, 634, 625

\bibitem[\protect\citeauthoryear{Robin et al.} {2003}]{robin2003}
   Robin, A. C., Reyl\'{e}, C., Derri\`{e}re, S. \& Picaud, S. 2003, A\&A, 409, 523.

\bibitem[\protect\citeauthoryear{Santos et al.}{2000}]{sb} 
    Santos, N. C., Israelian, G. \& Mayor, M., 2000,
    A \& A, 363, 228
\bibitem[\protect\citeauthoryear{Santos et al.}{2001}]{s1} 
    Santos, N. C., Israelian, G. \& Mayor, M., 2001,
    A \& A, 373, 1019

\bibitem[\protect\citeauthoryear{Santos et al.}{2004}]{s04} 
    Santos, N. C., Israelian, G. \& Mayor, M., 2004,
    A \& A, 415, 1153

\bibitem[\protect\citeauthoryear{Schiavon et~al.} {1997}]{Schiavon97}
   Schiavon, R.~P., Barbuy, B., \& Singh, P.~D., 1997,
   \newblock {  ApJ, }, 484, 499.

\bibitem[\protect\citeauthoryear{Tinney et al.}{2001}]{AAPSI}
  Tinney, C. G., Butler, R. P., Marcy, G. W., Jones, H. R. A., 
  Penny, A. J. Vogt, S. S., Apps, K. \& Henry, G. W.  2001, \apj, 551, 507

\bibitem[\protect\citeauthoryear{Tinney et al.}{2002a}]{AAPSIII}
  Tinney, C. G., Butler, R. P., Marcy, G. W., Jones, H. R. A., 
  Penny, A. J., McCarthy, C. \& Carter, B. D.  2002a, \apj, 571, 528

\bibitem[\protect\citeauthoryear{Tinney et al.}{2002b}]{CaHKI}
  Tinney, C. G., McCarthy, C., Jones, H. R. A., Butler, R. P., Carter, B. D.,
  Marcy, G. W.,  
  \& Penny, A. J.  2002b, \mnras, 332, 759

\bibitem[\protect\citeauthoryear{Tinney et al.}{2003}]{AAPSVII}
  Tinney, C. G., Butler, R. P., Marcy, G. W., Jones, H. R. A., Penny, A. J., McCarthy,
  C., Carter, B. D. \& Bond, J.  2003, ApJ, 587, 423

\bibitem[\protect\citeauthoryear{Tinney et al.}{2005}]{AAPSXI}
  Tinney, C. G., Butler, R. P., Marcy, G. W., Jones, H. R. A., Penny, A. J., McCarthy, C., 
  Carter, B. D. \& Fischer, D. A. 2005, ApJ, 623, 1171

\bibitem[\protect\citeauthoryear{Tinney et al.}{2006}]{AAPSXIII}
  Tinney, C. G., Butler, R. P., Marcy, G. W., Jones, H. R. A., Laughlin, G.,  
  Carter, B. D., Bailey, J. A. \& O'Toole, S. 2006, ApJ, 647, 594

\bibitem[\protect\citeauthoryear{Udry et al.} {2007}]{udry07}
   Udry, S., et al. 2007, A\&A, 469, L43


\bibitem[\protect\citeauthoryear{Valenti et~al.}{1995}]{Valenti95}
   Valenti, J.~A., Butler, R.~P. \& Marcy, G.~W. 1995,
   \newblock { PASP, } {107}, 966
   
\bibitem[\protect\citeauthoryear{Weis \& Upgren} {1995}]{wu95}
   Weis, E. W. \& Upgren, A. R. 1995, AJ, 109, 812

\bibitem[\protect\citeauthoryear{Wielen} {1977}]{wielen77}
  Wielen, R. 1977, A\&A, 60, 263

\end{thebibliography}
\end{document}